\documentclass[twocolumn,showpacs,preprintnumbers,amsmath,amssymb]{revtex4}
\usepackage{graphics, graphicx, float, amssymb}
\DeclareGraphicsExtensions{pdf}
\pdfoutput=1

\begin{document}

\title{Raman light scattering study and microstructural analysis\\
of epitaxial films of the electron-doped superconductor
La$_{2-x}$Ce$_x$CuO$_4$}
\author{M.~Rahlenbeck$^1$, M.~Wagenknecht$^{2,3}$, A.~Tsukada$^3$, D.~Koelle$^2$, R.~Kleiner$^2$, B.~Keimer$^1$, and C.~Ulrich$^{1,4,5}$}
\affiliation{$^1$Max-Planck-Institut f\"ur Festk\"orperforschung,
Heisenbergstra\ss e 1, D-70569 Stuttgart, Germany}
\affiliation{$^2$Physikalisches Institut - Experimentalphysik II,
Universit\"at T\"ubingen, Auf der Morgenstelle 14, D-72076
T\"ubingen, Germany}
\affiliation{$^3$NTT Basic Research
Laboratories, NTT Corporation, 3-1 Morinosato-Wakamiya, Atsugi,
Kanagawa 243-0198, Japan}
\affiliation{$^4$University of New South
Wales, School of Physics, 2052 Sydney, New South Wales, Australia}
\affiliation{$^5$The Bragg Institute, Australian Nuclear Science and
Technology Organization, Lucas Heights, NSW 2234, Australia}

\date{\today}

\begin{abstract}
We present a detailed temperature-dependent Raman light scattering
study of optical phonons in molecular-beam-epitaxy-grown films of
the electron-doped superconductor La$_{2-x}$Ce$_x$CuO$_4$ close to
optimal doping ($x\sim0.08$, $T_c = 29$ K and $x\sim 0.1$, $T_c =
27$ K). The main focus of this work is a detailed characterization
and microstructural analysis of the films. Based on micro-Raman
spectroscopy in combination with x-ray diffraction,
energy-dispersive x-ray analysis, and scanning electron microscopy,
some of the observed phonon modes can be attributed to micron-sized
inclusions of Cu$_2$O. In the slightly underdoped film
($x\sim0.08$), both the Cu$_2$O modes and others that can be
assigned to the La$_{2-x}$Ce$_x$CuO$_4$ matrix show pronounced
softening and narrowing upon cooling below $T \sim T_c$. Based on
control measurements on commercial Cu$_2$O powders and on a
comparison to prior Raman scattering studies of other
high-temperature superconductors, we speculate that proximity
effects at La$_{2-x}$Ce$_x$CuO$_4$/Cu$_2$O interfaces may be
responsible for these anomalies. Experiments on the slightly
overdoped La$_{2-x}$Ce$_x$CuO$_4$ film ($x\sim0.1$) did not reveal
comparable phonon anomalies.
\end{abstract}

\maketitle

Electron-doped high-temperature superconductors exhibit a
substantially lower maximum transition temperature and a narrower
doping range of superconductivity than their hole-doped
counterparts. The origin of this asymmetry of the phase diagram is
still incompletely understood, in part due to the complex materials
physics of the electron-doped cuprates. It has been demonstrated,
for instance, that the elaborate annealing procedure required to
stabilize superconductivity in bulk single crystals of
Nd$_{2-x}$Ce$_x$CuO$_4$ (NCCO), one of the most widely studied
electron-doped cuprates, generates a small amount of an epitaxially
intergrown impurity phase that profoundly affects the magnetic
properties.\cite{Greven04} Similar observations have been reported
for some epitaxial thin films of Pr$_{2-x}$Ce$_x$CuO$_4$ (PCCO)
grown by pulsed-laser deposition.\cite{Fournier09} The pervasiveness
of such inclusions and their influence on various physical
properties is only beginning to be explored, but the wide variation
of the transport characteristics of thin films of nominally
identical composition \cite{Greene} suggests that they may be quite
common.

Since Raman scattering is both a powerful spectroscopic probe of
superconductivity in the cuprates and an excellent diagnostic tool
for impurity phases, we have carried out a high-resolution Raman
scattering study of molecular beam epitaxy (MBE)-grown films of
La$_{2-x}$Ce$_x$CuO$_4$ (LCCO), the compound that exhibits both the
highest critical temperature and the widest doping range for
superconductivity among all electron-doped
cuprates.\cite{Naito00,Sawa02,Krockenberger08} LCCO crystalizes in
the so-called $T'$-structure, which does not include apical oxygen
ions, in contrast to the $T$-structure found in hole-doped members
of the $Ln_2$CuO$_4$ family. The stability of the $T'$-structure
depends on the radius of the lanthanide ($Ln$) ions.\cite{Naito00}
For $Ln$ = La, the $T'$-structure is unstable in bulk form, but can
be stabilized by epitaxial growth on SrTiO$_3$
substrates.\cite{Naito00} Depending on the growth conditions, LCCO
can also crystallize in the $T$-structure, but superconductivity is
observed only in the $T'$-structure.\cite{Naito02,Tsukada06} The
superconducting state is stable over a wider doping range ($0.05
\leq x \leq 0.22$) than in other electron-doped compounds such as
NCCO, and the maximum transition temperature ($T_c \sim 30$ K) is
found at a lower doping level ($x = 0.09$, compared to 0.15 for
NCCO).\cite{Tsukada05,Krockenberger08}

In the hole-doped cuprates, numerous Raman scattering experiments
have elucidated the magnitude, anisotropy, and doping dependence of
the superconducting energy gap, $\mathrm{\Delta}$, either directly
via electronic Raman scattering \cite{Hackl07} or indirectly via
phonon anomalies induced by the electron-phonon
interaction.\cite{Thomsen91,Bakr09} In electron-doped
superconductors, electronic Raman scattering
\cite{Blumberg02,Blumberg05} has also yielded valuable information
about the magnitude and anisotropy of $\mathrm{\Delta}$, but
superconductivity-induced phonon anomalies have thus far not been
reported. Since electronic Raman scattering is difficult in films,
we have used the latter method in an attempt to gain insight into
the energy gap of LCCO. We indeed observe phonon anomalies at
temperatures close to the superconducting transition temperature
$T_c$, but find that micron-sized Cu$_2$O inclusions present in all
samples complicate their interpretation.

The Raman scattering experiments were performed on two $900$ nm
thick LCCO films that had been the basis of prior transport
experiments.\cite{Wagenknecht08} The films had been deposited
epitaxially on [001]-oriented SrTiO$_3$ substrates, yielding c-axis
oriented films, in two independent fabrication runs using MBE from
pure metal sources, as described elsewhere.\cite{Naito00} After
growth, they were annealed in vacuum at 578$^\circ$C and
559$^\circ$C for 100 and 90 minutes, respectively, in order to
remove residual apical oxygen ions.\cite{Wagenknecht08a} The Ce
concentrations were adjusted to $x\sim 0.08$ (corresponding to
electron concentrations slightly less than optimum doping) and
$x\sim 0.1$ (slightly overdoped) by exact control of the gas flux
rates and use of inductively coupled plasma - atomic emission
spectroscopy (ICP-AES).\cite{Wagenknecht08a,Greibe01}
Electrical resistivity measurements revealed $T_c\sim29$ K and $27$
K, respectively. The widths of the resistivity transitions at $T_c$
were below 1 K for both films. For comparison, we used one
platelet-like Cu$_2$O single crystal with $c$-axis perpendicular to
the surface, and two commercial Cu$_2$O powder samples. Powder (I)
had a purity of $99.9\%$ with maximal impurity content of $0.006\%$
Fe, $0.004\%$ Si, $0.003\%$ Pb, $0.002\%$ Mn, and 3 ppm Sb (as
indicated by the supplier). Powder (II) had a purity of $99.5\%$,
and the supplier did not provide information about the nature of the
impurities.

For the Raman measurements we used two different setups. In order to
maximize the spectral resolution we used a ``macro" setup, where the
samples were mounted in a vertical helium-flow cryostat. The spectra
were taken in quasi-backscattering geometry using the linearly
polarized $514.5$ nm and $488.0$ nm lines of an Ar$^+$/Kr$^+$
mixed-gas laser for excitation. The laser beam was focused on a
$\sim100$ $\mu$m spot on the sample surface with an incident power
of less than $10$ mW, in order to avoid sample heating. The
scattered light was analyzed by a DILOR XY triple grating
spectrometer using a nitrogen-cooled charge-coupled-device (CCD)
camera. For higher spatial resolution we used a ``micro" setup,
where the samples were mounted on the cold finger of a horizontal
helium-flow cryostat. The spectra were taken in backscattering
geometry using the linearly polarized $532.0$ nm line of a frequency
doubled Nd:YAG laser for excitation. The laser beam was focused
through a $50\times$ ($10\times$) microscope objective to a $\sim3$
$\mu$m ($\sim10$ $\mu$m) diameter spot on the sample surface, with
an incident laser power of less than $1$ mW. The scattered light was
analyzed by a JobinYvon LabRam single grating spectrometer equipped
with a notch filter and a Peltier-cooled CCD camera. For each Raman
spectrum an additional calibration spectrum of a nearby argon or
neon line was measured in order to accurately determine the
frequency and linewidth of the different phonons. For data analysis,
all phonon peaks were fitted to Voigt profiles, which result from a
convolution of the Lorentzian phonon lineshape with the Gaussian
shaped instrumental resolution ($\sim4$ cm$^{-1}$ full width at half
maximum (FWHM) for the macro setup).

\begin{figure}
\includegraphics[width=8.5cm]{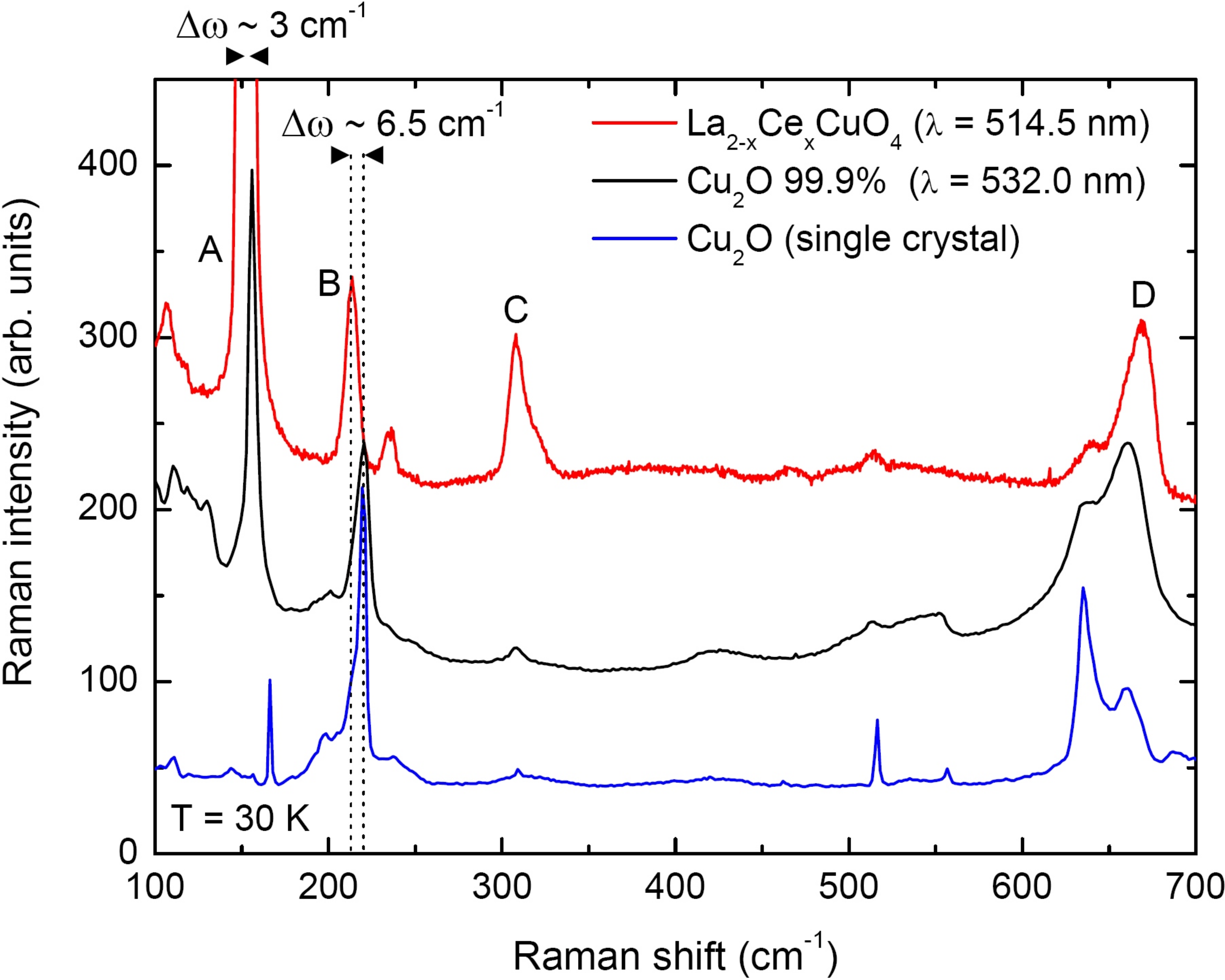}
\caption{\label{vergleich}Raman spectra at 30 K of (top) the
slightly underdoped LCCO film ($T_c\sim29$ K) in z(xx)z polarization
configuration for $514.5$ nm laser excitation, (middle) the Cu$_2$O
powder sample (99.9 \%), and (bottom) the Cu$_2$O single crystal
each for $532.0$ nm laser excitation. (Cu$_2$O spectra scaled for
clarity.)}
\end{figure}

Figure \ref{vergleich} shows the Raman spectrum of the slightly
underdoped LCCO film at temperature $T=30$ K in the z(xx)z
polarization configuration. Here, we use the Porto notation l(ij)m,
where l and m denote the direction of incident and scattered light
and (ij) their polarization, respectively. Both the slightly
underdoped and the slightly overdoped LCCO films show comparable
Raman spectra. We detect four main modes at 153.5 cm$^{-1}$ (mode
A), 213.5 cm$^{-1}$ (mode B), 308.5 cm$^{-1}$ (mode C), and a double
mode at $\sim640$-670 cm$^{-1}$ (mode D). A mode at $\sim 429$
cm$^{-1}$, which would correspond to an apical oxygen vibration in
the $T$-structure of La$_2$CuO$_4$ (Ref. \cite{Thomsen91}), is not
observed, indicating the absence of apical oxygen residuals. A group
theoretical analysis of the phonon modes of the $T'$-structure
(tetragonal space group $I4/mmm$ ($D_{4h}^{17}$)) \cite{Hadjiev89}
yields $A_{1g}+B_{1g}+2E_{g}$ Raman-active modes. To our knowledge,
Raman data on $T'$-LCCO have not yet been reported. We therefore use
data on related $T'$-lanthanide cuprates such as NCCO for comparison
\cite{Thomsen91,Sugai89,Heyen91}. The energies of modes B and C are
close to, but somewhat lower than those of the $A_{1g}$ and $B_{1g}$
vibrations, respectively, of the Nd and out-of-plane oxygen atoms in
$T'$-NCCO.\cite{Heyen91} A downward shift of the LCCO modes with
respect to NCCO would be in line with a continuous frequency evolution
previously observed in the series of lanthanide cuprates
\cite{Heyen91}.

\begin{figure}
\includegraphics[width=8.5cm]{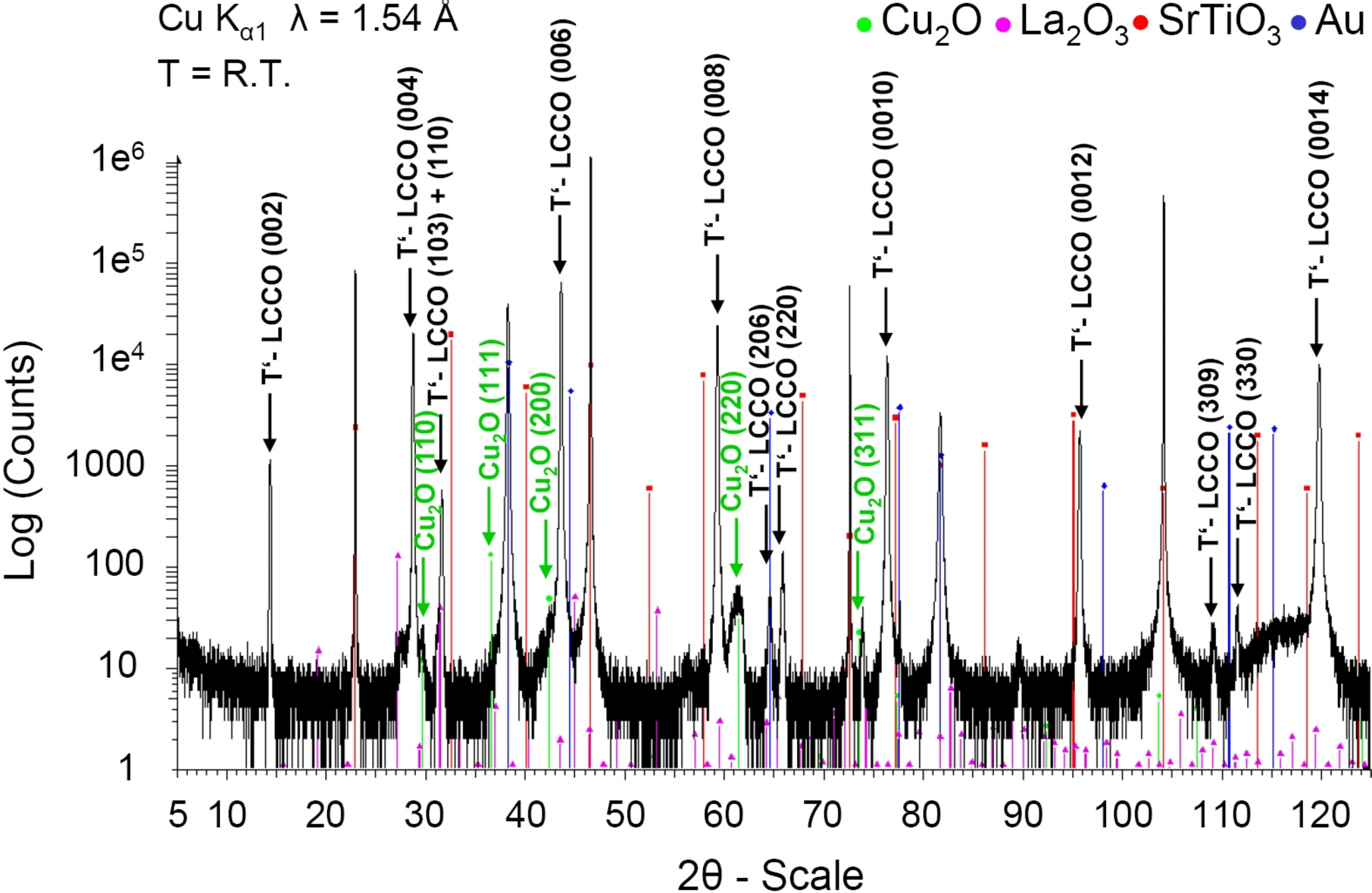}
\caption{\label{XRD}X-ray $\mathrm{\Theta}$-$\mathrm{2\Theta}$-scan
of the slightly underdoped LCCO film at room temperature (Cu
K$_{\alpha1}$ with $\lambda=1.54$ {\AA}). The Bragg reflections of
SrTiO$_3$ and gold are due to the substrate and gold contacts on top
of the film.}
\end{figure}

The assignment of modes A and D is more difficult, because Raman
spectra of other cuprates with the $T'$-structure do not exhibit
modes with similar energies. We have therefore investigated possible
contamination by impurity phases, including Cu$_2$O which has been
previously observed as a contaminant in NCCO.\cite{Hadjiev89} Figure
\ref{vergleich} shows a comparison to reference spectra of
compressed Cu$_2$O powder sample I (99.9 \% purity) and the Cu$_2$O
single crystal for 532.0 nm laser excitation. Based on this
comparison and on earlier data on Cu$_2$O
\cite{Petroff75,Petroff72}, modes A and D can be identified with the
infrared-allowed $\Gamma_{15}^{(1)}$ and $\Gamma_{15}^{(2)}$ modes
of Cu$_2$O.\cite{Huang63} Due to resonance effects, these modes can
become Raman-active with high intensity.\cite{Petroff74,Yu73} The
much stronger intensity of the $\Gamma_{15}^{(1)}$ mode in the
Cu$_2$O powder compared to the Cu$_2$O single crystal can be
attributed to powder averaging over all possible polarization
geometries. Mode A in the LCCO films also shows high intensity,
suggesting an isotropic orientation of the Cu$_2$O impurity phase.
The shift of $\sim3$ cm$^{-1}$ of mode A with respect to the
$\Gamma_{15}^{(1)}$ mode of Cu$_2$O may be a consequence of stress
imposed by the LCCO matrix. Note that modes with energies roughly
comparable to those of modes B and C are also present in the Cu$_2$O
reference spectra, but the former mode is shifted by $\sim 6.5$
cm$^{-1}$ with respect to mode B, and the latter mode is extremely
weak. We will show below that these modes likely originate from
LCCO.

In order to directly characterize the chemical composition of the
LCCO films, we used high-intensity x-ray diffraction (XRD). Figure
\ref{XRD} shows a $\mathrm{\Theta}$-$\mathrm{2\Theta}$ scan of the
slightly underdoped LCCO film at room temperature, plotted on a
logarithmic intensity scale. Based on a comparison with the
calculated Bragg angles for the $T'$-structure \cite{Naito00}, we
clearly identify the expected [001]-oriented $T'$-structure as the
main phase. Additional XRD pole figure measurements (not shown here)
also reveal the Bragg peaks (103) and (110). This indicates two
$T'$-LCCO minority phases with different growth directions, which
are stabilized by a well-known, accidental match between the
in-plane and out-of-plane lattice parameters ($c/a\approx3$). We
find no indication for the presence of any $T$-structure inclusions.
The XRD measurements also confirm the presence of a Cu$_2$O impurity
phase, as well as a trace amount of La$_2$O$_3$ (Fig. \ref{XRD}).
The intensities of the main Cu$_2$O Bragg reflections are about four
orders of magnitude below those of LCCO, but their ratios indicate
random orientation of the Cu$_2$O crystallites, in contrast to the
epitaxially oriented LCCO matrix. Taking powder averaging of the
Cu$_2$O Bragg reflections into account, the XRD data imply that the
volume fraction of Cu$_2$O is only about an order of magnitude below
the one of LCCO.

\begin{figure}
\includegraphics[width=8.0cm]{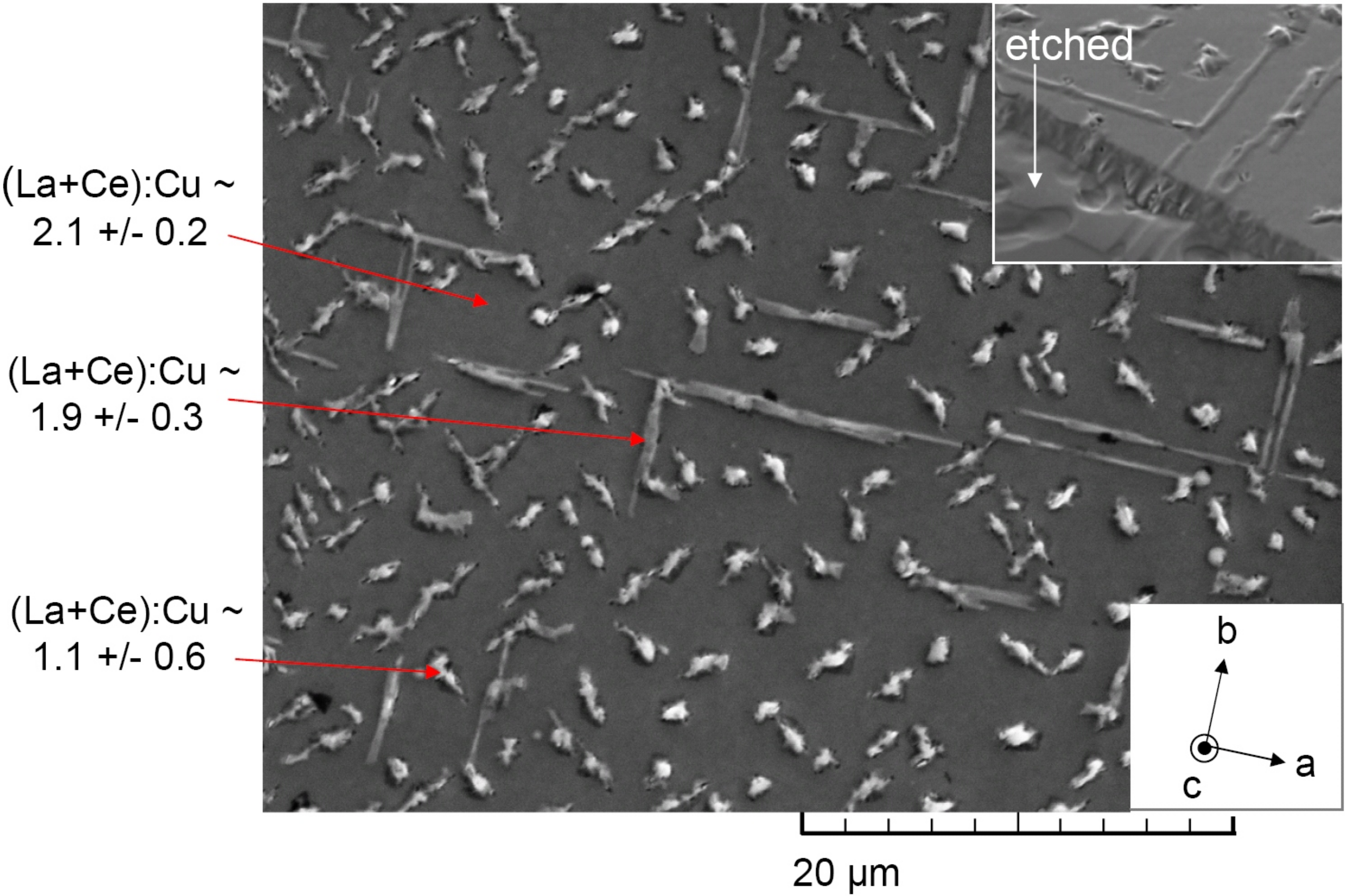}
\caption{\label{surface}SEM image of the slightly underdoped LCCO
film. The upper inset shows a partly etched surface area.}
\end{figure}

The microstructure and the local variation of the chemical
composition of the LCCO films were further analyzed by scanning
electron microscopy (SEM) and energy dispersive x-ray (EDX)
analysis. The microstructure of both LCCO films appeared comparable.
Figure \ref{surface} shows a SEM image of the slightly underdoped
LCCO film. About 18\% of the surface area is uniformly flecked with
particles of $\sim 3$ $\mu$m diameter. While the surface background
exhibits the composition ratio (La+Ce):Cu $\sim 2.1 \pm 0.2$
characteristic of $T'$-LCCO, the particles were found to be centers
of strongly enhanced Cu content, with the statistically averaged
ratio (La+Ce):Cu $\sim 1.1 \pm 0.6$. They are thus good candidates
for the Cu$_2$O impurity phase. The high and uniform surface
coverage with particles supports the picture of an isotropically
oriented Cu$_2$O impurity phase of $\gtrsim10\%$ in the LCCO films.
In addition, we observed line structures oriented along the $ab$
crystal axes, which exhibit a slightly enhanced Cu content. These
structures may originate from structural defects or from the
$T'$-LCCO minority phases. In the inset of figure \ref{surface} we
show a surface area of the slightly underdoped LCCO film, which was
partly etched by Ar ion bombardment. We find that both the particles
and the line structures reach deeply into the film, suggesting their
formation during the crystal growth. The concentrations of impurity
atoms beyond the constituent elements of LCCO were below the EDX
detection limit.

\begin{figure}[t]
\includegraphics[width=8.4cm]{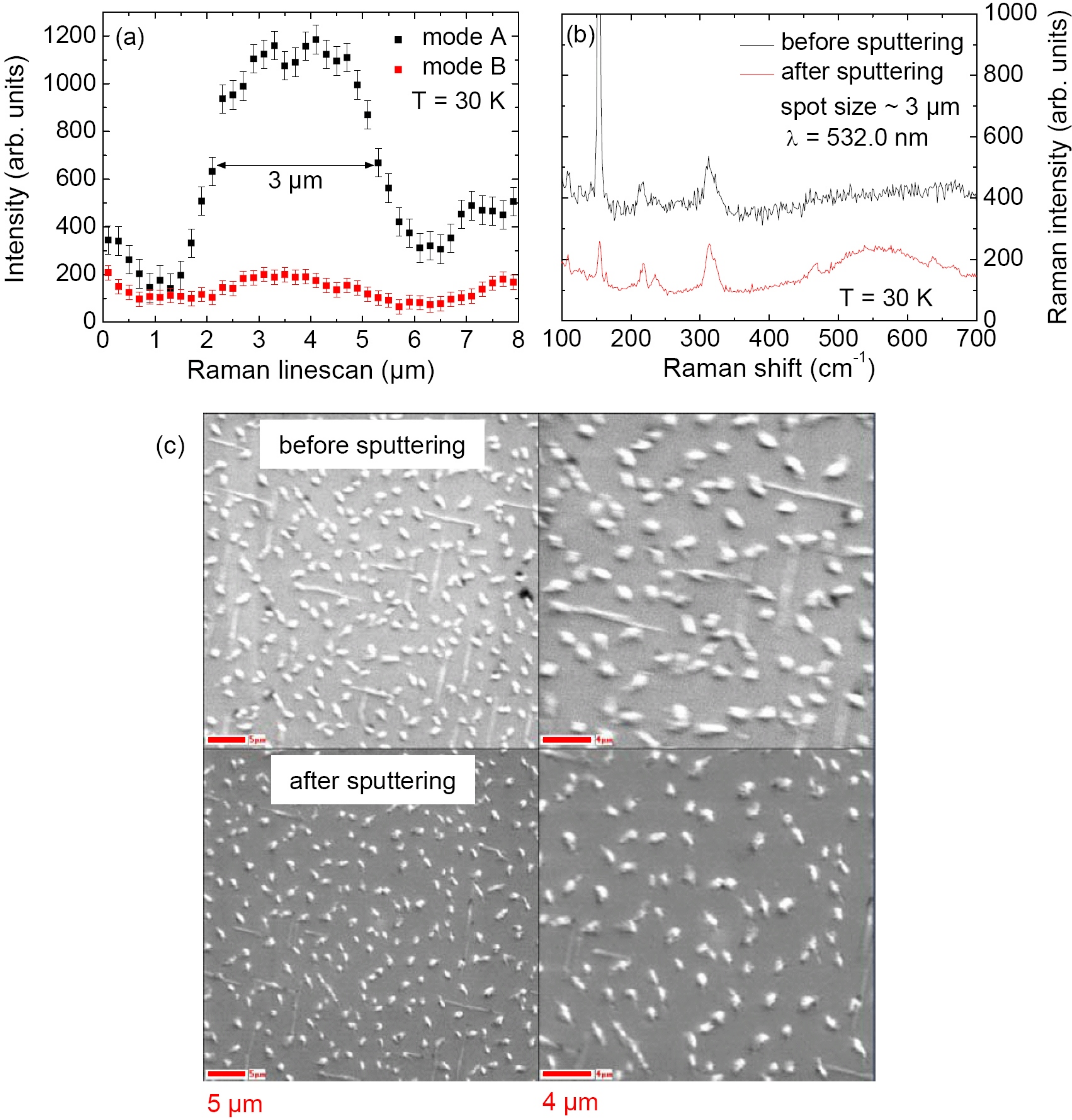}
\caption{\label{sputtering}(a) Line scan of the intensities of modes
A and B of the underdoped LCCO film taken in the micro-Raman setup
with $50\times$ microscope objective and $\lambda=532.0$ nm, (b)
Raman spectra at $T=30$ K before and after sputtering of $\sim250$
nm, and (c) SEM images before and after sputtering.}
\end{figure}

In order to relate the Raman spectra of Fig. \ref{vergleich} to the
microscopic observations of Fig. \ref{surface}, we employed the
micro-Raman setup. Figure \ref{sputtering}a shows a line scan of the
intensity of mode A in the slightly underdoped LCCO film. The
intensity varies strongly on a length scale of $\sim3$ $\mu$m, which
is comparable to the size of the particles in Fig. \ref{surface}.
This underscores the assignment of this mode to the Cu$_2$O impurity
phase. The intensities of modes B and C, on the other hand, depend
only weakly on the measuring position (Fig. \ref{sputtering}a),
supporting the conclusion that they do not
originate from Cu$_2$O, but from the $T'$-LCCO host material. We
used argon ion sputtering under vacuum in an attempt to remove the
Cu$_2$O particles. A comparison of SEM images of the surface before
and after removal of $\sim250$ nm (Fig. \ref{sputtering}c) confirms
that the particles are stuck deeply inside the LCCO matrix. While it
is thus not possible to remove the particles, sputtering still
reduces both the volume fraction of Cu$_2$O and the intensity of
mode A (Fig. \ref{sputtering}b). Modes B and C, on the other hand,
are nearly unaffected by sputtering.

\begin{figure}
\centering
\includegraphics[width=7.5cm]{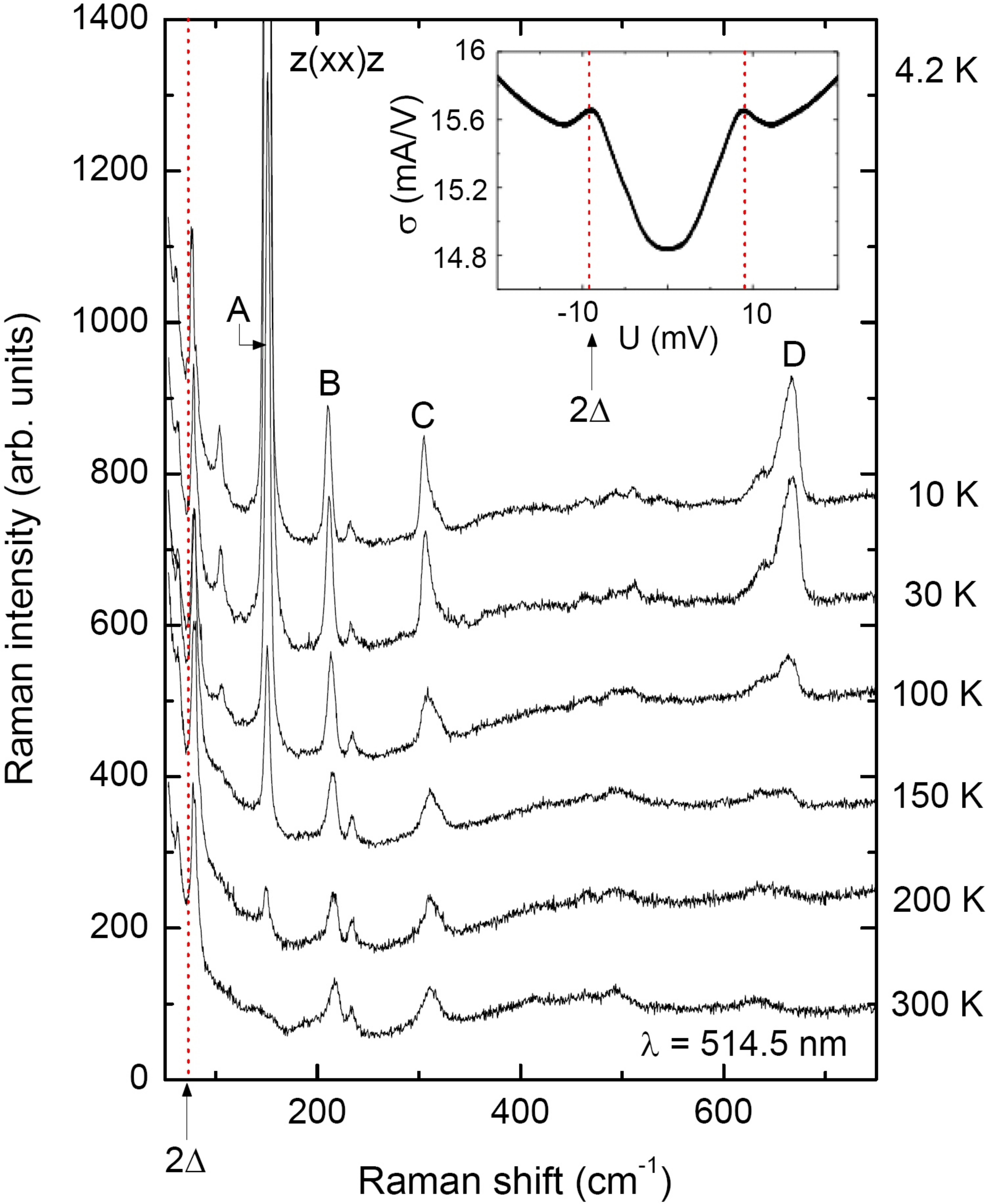}
\caption{\label{spectrum}Raman spectra of the slightly underdoped
LCCO film at temperatures 10 K $\leq T \leq 300$ K in z(xx)z
polarization configuration for $514.5$ nm laser excitation.
$2\mathrm{\Delta}\sim9$ meV denotes the energy of the
superconducting gap. The inset shows the corresponding quasiparticle
conductance at $4.2$ K.}
\end{figure}

\begin{figure*}
\centering
\includegraphics[width=14.5cm]{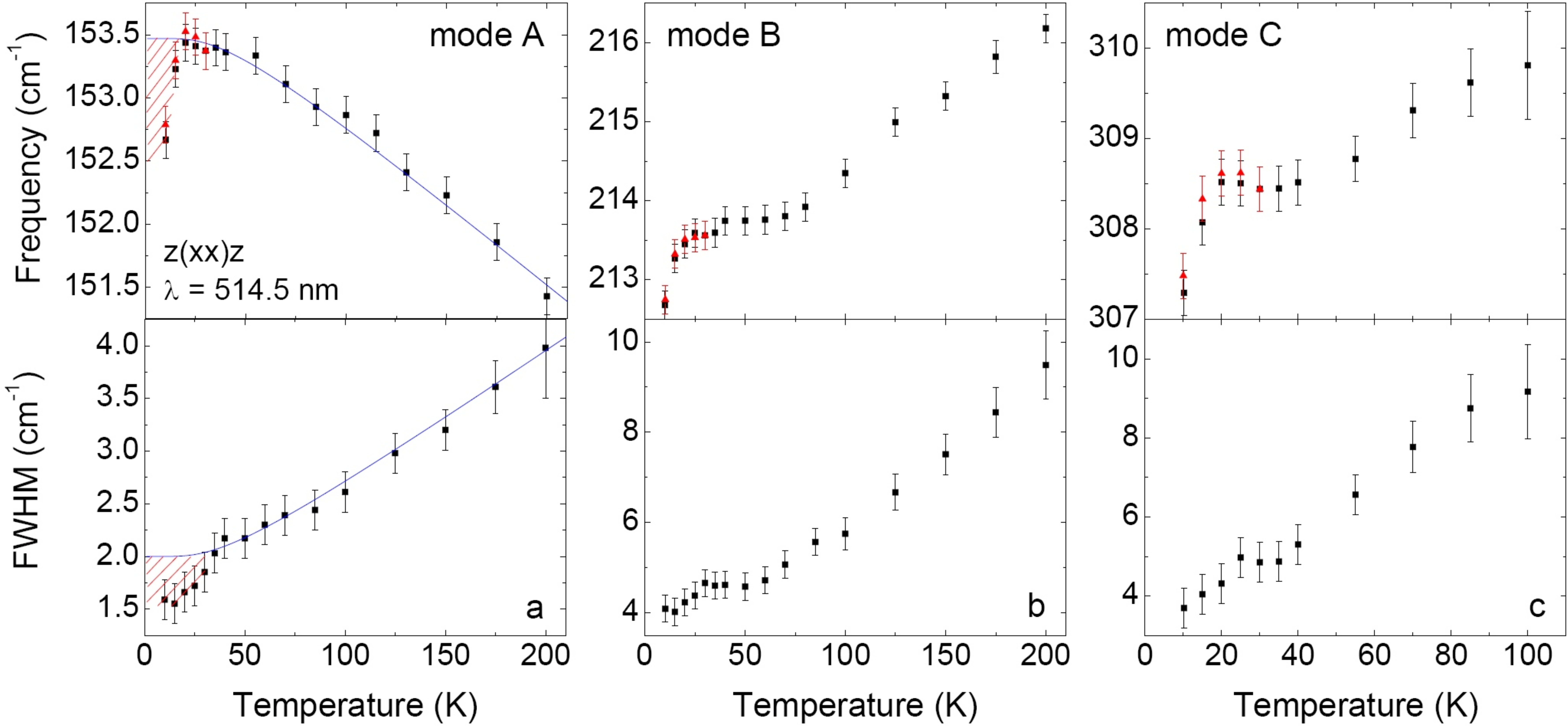}
\caption{\label{anomalies}Temperature dependence of the frequency
and FWHM of modes A, B, and C in the underdoped LCCO film (see Fig.
\ref{vergleich}). The triangular points represent a second
independent data set at low temperatures. The solid lines are the
result of fits to the data points above $T_c$ according to the
theory of anharmonic phonon decay \cite{Menendez84,Hadjiev98} (see
text for details). The shaded areas indicate deviations from the
anharmonic temperature dependence below $\sim T_c$.}
\end{figure*}

Having obtained a thorough understanding of the microstructure and
phase composition of the LCCO films, we now focus on anomalies in
the temperature dependence of the different optical modes at the
superconducting transition temperature $T_c$. Figure \ref{spectrum}
shows the Raman spectra of the slightly underdoped LCCO film from
$10$ K to $300$ K in z(xx)z polarization for $514.5$ nm laser
excitation. The integrated intensities of modes A and D show a
strong temperature dependence with a significant increase below
$T\sim150$ K. A similar activation was also observed in the Cu$_2$O
reference samples (not shown here),
again confirming their common origin. In contrast, the intensities
of modes B and C exhibit a much weaker temperature dependence.
Figure \ref{anomalies}a shows the temperature dependence of the
frequency and FWHM of mode A. The solid lines are the result of fits
to the data above $T_c$, using an expression based on anharmonic
phonon-phonon interactions \cite{Menendez84,Hadjiev98}. For
simplicity we assumed a symmetric decay into two product modes,
which leads to the following expressions for the phonon frequency
$\omega_{ph}$ and FWHM $\Gamma_{ph}$:
\begin{eqnarray*}
\omega_{ph}(T)=-A\left(1+\frac{2a}{exp(\hbar\omega_{0}/2k_BT)-1}\right)+\omega_0,\\
\Gamma_{ph}(T)=\Gamma_{a}\left(1+\frac{2a}{exp(\hbar\omega_{0}/2k_BT)-1}\right)+\Gamma_b,
\end{eqnarray*}
where $A$ and $\Gamma_{a}$ are positive constants and $a$ corrects
for terms arising from nonsymmetric phonon decay processes.
$\Gamma_b$ represents the temperature-independent part of
$\Gamma_{ph}$.

\begin{figure}
\includegraphics[width=8.5cm]{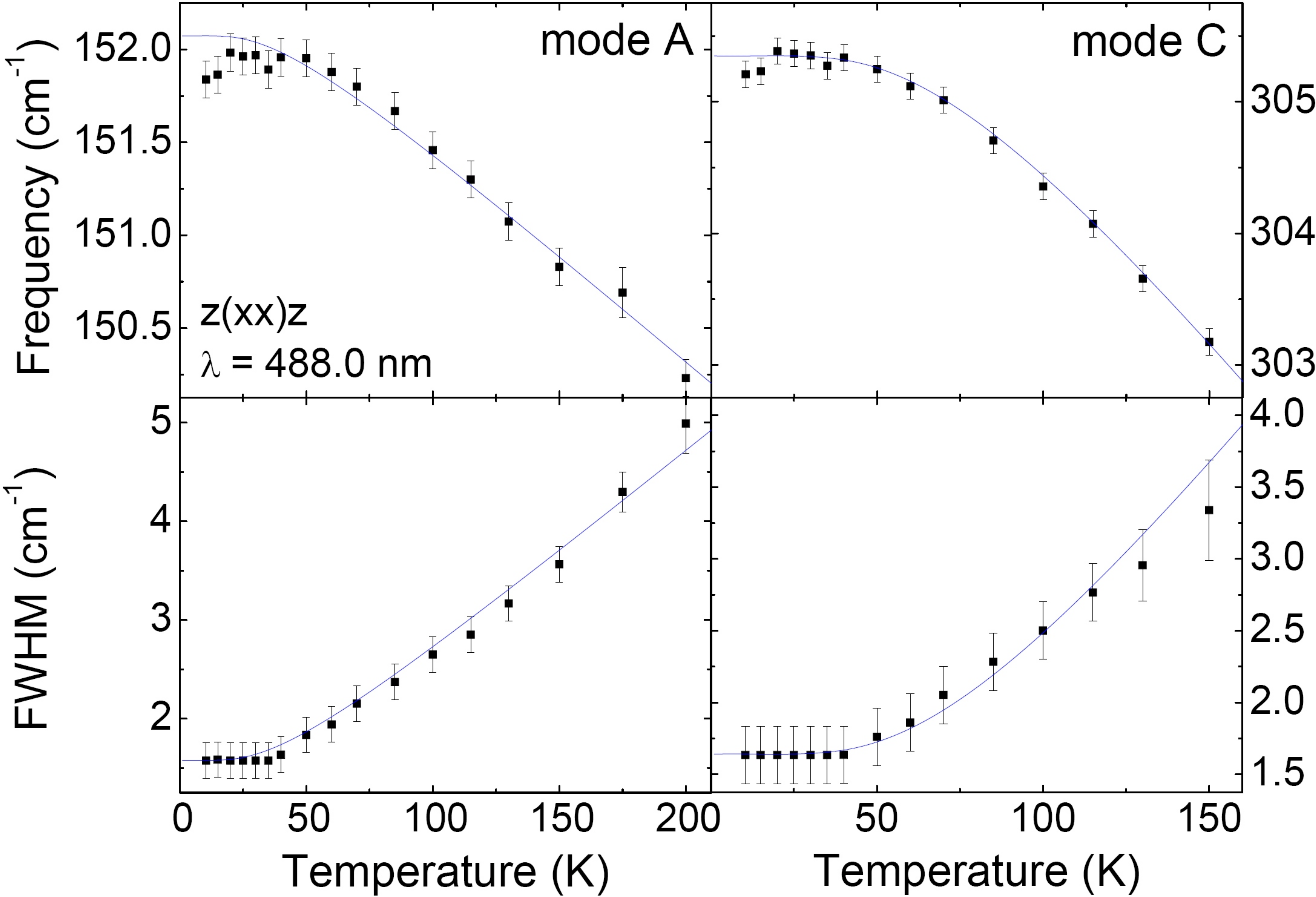}
\caption{\label{overdoped}Temperature dependence of the frequency
and FWHM of modes A and C in the slightly overdoped LCCO film
($T_c\sim27$ K). The $487.986$ nm laser line was used for
excitation. For solid lines see caption of figure \ref{anomalies}.}
\end{figure}

While mode A follows nearly perfectly the expression for anharmonic
decay in the normal state, we observe significant deviations from
this behavior below $T \sim T_c$. The frequency softens by $\sim0.7$
cm$^{-1}$ upon cooling below $T_c$, and the temperature dependence
of the linewidth exhibits a change in slope in the same temperature
range, which corresponds to a narrowing of $\sim0.5$ cm$^{-1}$. The
deviations from the anharmonic behavior are illustrated by the
shaded areas in figure \ref{anomalies}a. In contrast to mode A, the
temperature dependence of the parameters characterizing modes B and
C differs substantially from the standard anharmonic behavior (Figs.
\ref{anomalies}b and c). In particular, their frequencies increase
continuously with increasing $T$, opposite to the behavior expected
from anharmonicity. Below $T_c$, however, they exhibit softening and
narrowing of the same magnitude as the one observed for mode A.
Remarkably, both the anomalous normal-state behavior of mode C and
the renormalization of the phonon frequencies below $T_c$ are absent
in the slightly overdoped LCCO film (Fig. \ref{overdoped}),
suggesting that these features are controlled by the doping level of
LCCO.

\begin{figure}[h]
\includegraphics[width=8.8cm]{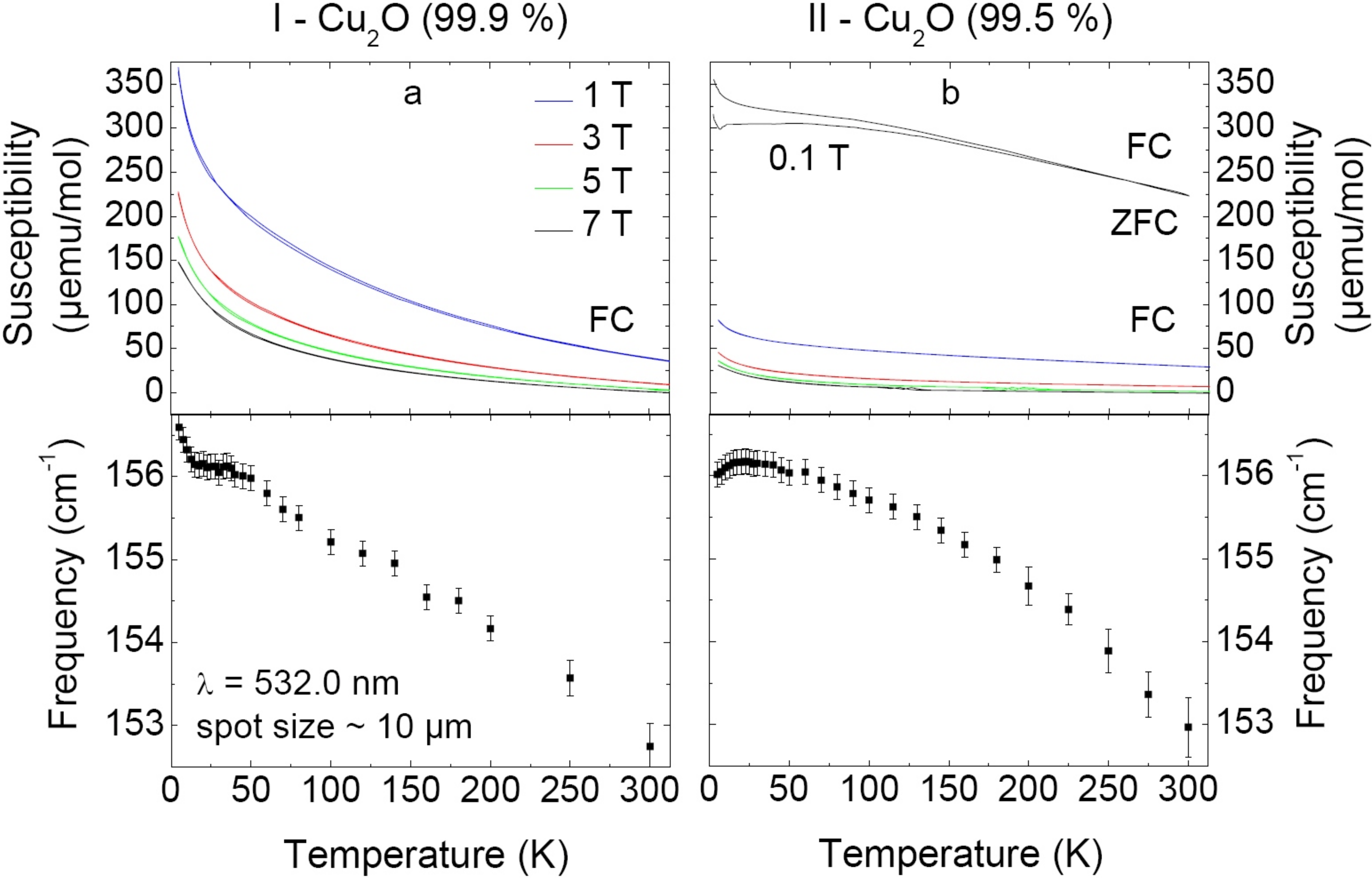}
\caption{\label{Cu2O}Temperature dependence of the magnetic
susceptibility for different magnetic fields (FC = field cooled, ZFC
= zero field cooled), and frequency of the $\Gamma_{15}^{(1)}$ mode
of two compressed Cu$_2$O powder samples with different purity
levels.}
\end{figure}

For comparison we have measured the temperature dependence of the
magnetic susceptibility and phonon frequencies of the two commercial
Cu$_2$O powders. Although pure Cu$_2$O is
nonmagnetic,\cite{Schlafer69} the susceptibilities of both powders
exhibit pronounced low-temperature Curie tails due to magnetic
impurities (Fig. \ref{Cu2O}). The susceptibility of one of the two
samples even exhibits a small anomaly at low magnetic fields and $T
\sim 6$ K, and some field hysteresis over a wider temperature range
(Fig. \ref{Cu2O}b). These observations agree qualitatively with the
recent report of a substantial influence of cation vacancies and
small amounts of magnetic impurities on the magnetic properties of
Cu$_2$O.\cite{Chen09} The overall temperature dependence of the
frequency of the intense $\Gamma_{15}^{(1)}$ mode shown in Fig.
\ref{Cu2O} is consistent with the one expected for anharmonic decay.
In powder sample I,
however, the mode abruptly hardens by $\sim0.5$ cm$^{-1}$ between $T
\sim 15$ K and the base temperature of 5 K (Fig. \ref{Cu2O}a), while
powder sample II shows a small hint of a softening (Fig.
\ref{Cu2O}b). These phonon anomalies are surprising, because
low-temperature structural instabilities have not been reported for
Cu$_2$O.\cite{Werner82} Moreover, they are apparently uncorrelated
with features in the magnetic susceptibility.
The phonon anomaly therefore likely arises from sample-specific
defects or impurities. Note that the anomaly of the Cu$_2$O
vibration in powder sample I with higher purity level is smaller in
magnitude and of opposite sign than the one exhibited by mode A in
the underdoped LCCO film (which was fabricated from ultrapure metal
sources and is therefore much less affected by magnetic impurities
than the commercial powders), and that it occurs at a lower
temperature. Nonetheless, the Raman data on Cu$_2$O do indicate that
the low-temperature behavior of the $\Gamma_{15}^{(1)}$ mode is
quite sensitive to microstructural details. This may provide clues
to the origin of the anomalous low-temperature behavior of mode A in
the LCCO film, which we had identified with the $\Gamma_{15}^{(1)}$
mode of the Cu$_2$O inclusions.

In summary, all of the Raman-active phonons observed in the
underdoped LCCO film (including modes B and C that are likely due to
the LCCO host material as well as mode A, which likely arises from
Cu$_2$O inclusions) exhibit anomalies in their temperature
dependence that are reproducible and clearly outside the
experimental error bars at a temperature that is consistent with
$T_c$. In discussing these observations, we first ignore the Cu$_2$O
inclusions and consider the standard picture of
superconductivity-induced phonon anomalies which has been
established based on Raman data on hole-doped high-temperature
superconductors.\cite{Zeyher90,Thomsen91,Devereaux94} According to
this theory, optical phonons with energies higher than twice the
superconducting gap, $2\mathrm{\Delta}$, harden below $T_c$, and
their linewidths increase due to an enhanced density of states above
the gap. Conversely, phonons with energies below $2\mathrm{\Delta}$
are expected to soften.
Deviations from this behavior are predicted only close to
$2\mathrm{\Delta}$.\cite{Zeyher90} A microscopic formulation of this
theory yields a quantitative description of electronic Raman
scattering and superconductivity-induced phonon self-energy
anomalies in hole-doped YBa$_2$Cu$_3$O$_{6+y}$.\cite{Bakr09} The
superconducting energy gap LCCO is known from prior transport
measurements on the same films that we have investigated by Raman
light scattering.\cite{Wagenknecht08} Tunneling characteristics (one
of which is reproduced in the inset of Fig. \ref{spectrum}) show
coherence peaks at an energy of 9 meV, which implies
$2\mathrm{\Delta}\sim75$ cm$^{-1}$. Since the energies of all of the
Raman-active phonons we have discussed are far above 75 cm$^{-1}$,
the standard model predicts a weak hardening and broadening below
$T_c$, in complete contrast to our observations.

Since the standard theory of superconductivity-induced phonon
renormalization fails to account for the observations displayed in
Fig. \ref{anomalies}, we are forced to consider more unconventional
scenarios. A possible explanation of the softening and narrowing of
the optical modes below $T_c$ would be a second gap with magnitude
in excess of $2\mathrm{\Delta^*}\sim100$ meV that opens at a
temperature close to $T_c$ or is at least affected by the
superconducting phase transition. In this case, all of the
Raman-active optical modes would be located below this threshold
energy, and the observed softening could be explained by a
straightforward application of the theory of phonon self-energy to
this high-energy gap. The narrowing of the phonon linewidths would
then be a direct consequence of the reduced number of relaxation
channels due to the loss of spectral weight below
$2\mathrm{\Delta^*}$. Angle-resolved photoemission spectroscopy
\cite{Armitage01b,Matsui07} and optical spectroscopy
\cite{Onose01,Zimmers05} experiments on underdoped NCCO have indeed
yielded evidence of a high-energy ``pseudogap" that opens up below a
temperature $T^* > T_c$. Both $T^*$ and the magnitude of
$\mathrm{\Delta^*}$ were found to decrease with increasing doping
level, \cite{Matsui07,Dagan05}, so that $T^*$ cuts
the superconducting phase boundary at optimal doping.\cite{Zimmers05}
In the superconducting regime of NCCO, $\mathrm{\Delta^*}\sim 100$
meV $\gg \Delta$.\cite{Matsui07} Although no evidence has yet been
reported of a similar phenomenon in LCCO, it is thus conceivable
that interplay between superconductivity and the pseudogap (which
may in turn be related to the presence of antiferromagnetic order
\cite{Matsui07}) could explain the unusual superconductivity-induced
phonon self-energy anomalies we have observed. However, nearly
identical anomalies exhibited by mode A (which, as we have argued,
probably originates in the Cu$_2$O inclusions) point to a more
complex picture in which proximity and/or inverse proximity effects
at the LCCO/Cu$_2$O interface are also involved. It is possible, for
instance, that charge transfer across the interface leads to the
formation of magnetic Cu$^{2+}$ ions and induces cooperative
magnetism at the boundaries of the Cu$_2$O inclusions, which in turn
enhances antiferromagnetic order and the pseudogap in LCCO. An
investigation of phonon anomalies close the interface by Raman
spectroscopy with spatial resolution comparable to the
superconducting coherence length is well beyond our current
experimental capabilities, but may become possible in the future
based on advances in near-field optics. Based on our observations,
it appears worthwhile to study such effects systematically using
other experimental methods.

In addition to its scientific interest, the technical aspects of our
study are of general relevance for the investigation of
electron-doped high-temperature superconductors. Inclusions of
impurity phases such as lanthanide oxides \cite{Greven04,Fournier09}
and Cu$_2$O are hard to avoid during synthesis and continue to be
present in state-of-the-art crystals and films. They are also
difficult to detect based on standard x-ray diffraction, either
because epitaxial intergrowth leads to Bragg peak positions that are
similar to those of the host phase \cite{Greven04,Fournier09} or
because powder averaging greatly reduces the intensity of the
impurity Bragg reflections. Note, in particular, that the Cu$_2$O
inclusions in our LCCO films required an experimental setup with a
ratio of LCCO Bragg intensities to the noise floor of $\sim 10^4$
(Fig. \ref{XRD}), which goes beyond the typical diagnostics run on
thin films. We have shown that both electron microscopy and
micro-Raman spectroscopy are powerful, complementary diagnostic
tools (Fig. \ref{sputtering}). As we have seen, a thorough
understanding of impurity inclusions is important not only for
transport, but also for spectroscopic measurements.

We thank Y. Kuru, B. Bohnenbuck, C. Busch, M. Konuma, E. Br\"ucher,
M. Schaloske, R.K. Kremer, and A. Schulz for experimental support,
and M. Cardona, D. Manske, and R. Zeyher for fruitful discussions.
This work was supported by the Deutsche Forschungsgemeinschaft (DFG)
through project Kl930/11.


\begin{thebibliography}{}

\bibitem{Greven04} P.K. Mang, S. Larochelle, A. Mehta, O.P. Vajk, A.S. Erickson, L. Lu, W.J.L. Buyers, A.F. Marshall, K. Prokes, and M. Greven, Phys. Rev. B {\textbf 70}, 094507 (2004)
\bibitem{Fournier09} G. Roberge, S. Charpentier, S. Godin-Proulx, P. Rauwel, K.D. Truong, P. Fournier, J. Cryst. Growth {\textbf 311}, 1340 (2009)
\bibitem{Greene} For a review, see N.P. Armitage, P. Fournier, and R.L. Greene, preprint arXiv:0906.2931
\bibitem{Naito00} M. Naito and M. Hepp, Jpn. J. Appl. Phys. \textbf{39}, L485 (2000)
\bibitem{Sawa02} A. Sawa, M. Kawasaki, H. Takagi, Y. Tokura, Phys. Rev. B {\textbf 66}, 014531 (2002)
\bibitem{Krockenberger08} Y. Krockenberger, J. Kurian, A. Winkler, A. Tsukada, M. Naito, and L. Alff, Phys. Rev. B \textbf{77}, 060505(R) (2008)
\bibitem{Naito02} M. Naito, A. Tsukada, T. Greibe, and H. Sato, Proc. SPIE \textbf{4811}, 140 (2002)
\bibitem{Tsukada06} A. Tsukada, H. Yamamoto, and M. Naito, Phys. Rev. B \textbf{74}, 174515 (2006)
\bibitem{Tsukada05} A. Tsukada, Y. Krockenberger, M. Noda, H. Yamamoto, D. Manske, L. Alff, and M. Naito, Solid State Commun. \textbf{133}, 427 (2005)
\bibitem{Hackl07} For a review, see T.P. Devereaux and R. Hackl, Rev. Mod. Phys. {\textbf 79}, 175 (2007)
\bibitem{Thomsen91} C. Thomsen, in \textit{Light Scattering in Solids VI, Topics Appl. Phys.}, edited by M. Cardona and G. G\"{u}ntherodt (Springer, 1991), Vol. 68, p. 285
\bibitem{Bakr09} M. Bakr, A.P. Schnyder, L. Klam, D. Manske, C.T. Lin, B. Keimer, M. Cardona, and C. Ulrich, Phys. Rev. B \textbf{80}, 064505 (2009)
\bibitem{Blumberg02} G. Blumberg, A. Koitzsch, A. Gozar, B.S. Dennis, C.A. Kendziora, P. Fournier, and R.L. Green, Phys. Rev. Lett. \textbf{88}, 107002 (2002)
\bibitem{Blumberg05} M.M. Qazilbash, A. Koitzsch, B.S. Dennis, A. Gozar, Hamza Balci, C.A. Kendziora, R.L. Greene, and G. Blumberg, Phys. Rev. B {\textbf 72}, 214510 (2005)
\bibitem{Wagenknecht08} M. Wagenknecht, D. Koelle, R. Kleiner, S. Graser, N. Schopohl, B. Chesca, A. Tsukada, S.T.B. Goennenwein, and R. Gross, Phys. Rev. Lett. \textbf{100}, 227001 (2008); M. Wagenknecht, M. Rahlenbeck, D. Koelle, R. Kleiner, A. Tsukada, S.T.B. Goennenwein, and R. Gross, submitted to Phys. Rev. B; M. Rahlenbeck, unpublished data
\bibitem{Wagenknecht08a} M. Wagenknecht, \textit{Korngrenz-Tunnelspektroskopie am elektronendotierten Kupratsupraleiter La$_{2-x}$Ce$_x$CuO$_4$}, Ph.D. thesis, T\"ubingen University, 2008
\bibitem{Greibe01} T. Greibe, \textit{MBE growth of superconducting thin films}, Technical report, NTT Basic Research Laboratories, 2001
\bibitem{Hadjiev89} V.G. Hadjiev, I.Z. Kostadinov, L. Bozukov, E. Dinolova, and D.M. Mateev, Solid State Commun. \textbf{71}, 1093 (1989)
\bibitem{Sugai89} S. Sugai, T. Kobayashi, and J. Akimitsu, Phys. Rev. B \textbf{40}, 2686 (1989)
\bibitem{Heyen91} E.T. Heyen, R. Liu, M. Cardona, S. Pi\~{n}ol, R.J. Melville, D. McK. Paul, E. Mor\'{a}n, and M.A. Alario-Franco, Phys. Rev. B \textbf{43}, 2857 (1991)
\bibitem{Petroff75} Y. Petroff, P.Y. Yu, and Y.R. Shen, Phys. Rev. B \textbf{12}, 2488 (1975)
\bibitem{Petroff72} Y. Petroff, P.Y. Yu, and Y.R. Shen, Phys. Rev. Lett. \textbf{29}, 1558 (1972)
\bibitem{Huang63} K. Huang, Z. Phys. {\bf 171}, 213 (1963)
\bibitem{Petroff74} Y. Petroff, J. Phys. (Paris) Colloq. C3, \textbf{35}, 277 (1974)
\bibitem{Yu73} P.Y. Yu, Y.R. Shen, and Y. Petroff, Solid State Commun. \textbf{12}, 973 (1973)
\bibitem{Menendez84} J. Men\'{e}ndez and M. Cardona, Phys. Rev. B \textbf{29}, 2051 (1984)
\bibitem{Hadjiev98} V.G. Hadjiev, X. Zhou, T. Strohm, and M. Cardona, Phys. Rev. B \textbf{58}, 1043 (1998)
\bibitem{Schlafer69} H.L. Schlafer and G. Gliemann, \textit{Basic principles of Ligand Field Theory} (Wiley, New York 1969) p. 120
\bibitem{Chen09} Ch. Chen, L. He, L. Lai, H. Zhang, J. Lu, L. Guo, and Y. Li, J. Phys.: Condens. Matter \textbf{21}, 145601 (2009)
\bibitem{Werner82} A. Werner and H. D. Hochheimer, Phys. Rev. B \textbf{25}, 5929 (1982)
\bibitem{Zeyher90} R. Zeyher and G. Zwicknagl, Z. Phys. B - Condensed Matter \textbf{78}, 175 (1990)
\bibitem{Devereaux94} T.P. Devereaux, Phys. Rev. B \textbf{50}, 10287 (1994)
\bibitem{Armitage01b} N.P. Armitage, D.H. Lu, C. Kim, A. Damascelli, K.M. Shen, F. Ronning, D.L. Feng, P. Bogdanov, and Z.-X. Shen, Phys. Rev. Lett. \textbf{87}, 147003 (2001)
\bibitem{Matsui07} H. Matsui, T. Takahashi, T. Sato, K. Terashima, H. Ding, T. Uefuji, and K. Yamada, Phys. Rev. B \textbf{75}, 224514 (2007)
\bibitem{Onose01} Y. Onose, Y. Taguchi, K. Ishizaka, and Y. Tokura, Phys. Rev. Lett. \textbf{87}, 217001 (2001)
\bibitem{Zimmers05} A. Zimmers, J.M. Tomczak, R.P.S.M. Lobo, N. Bontemps, C.P. Hill, M.C. Barr, Y. Dagan, R.L. Greene, A.J. Millis, and C.C. Homes, Europhys. Lett. \textbf{70}, 225 (2005)
\bibitem{Dagan05} Y. Dagan, M.M. Qazilbash, and R.L. Greene, Phys. Rev. Lett. {\textbf 94}, 187003 (2005)

\end{thebibliography}
\end{document}